# An Efficient Dynamic and Distributed RSA Accumulator

Michael T. Goodrich[1]   Roberto Tamassia[2]   Jasminka Hasić[2]

**Abstract**

We show how to use the RSA one-way accumulator to realize an efficient and dynamic authenticated dictionary, where untrusted directories provide cryptographically verifiable answers to membership queries on a set maintained by a trusted source. Our accumulator-based scheme for authenticated dictionaries supports efficient incremental updates of the underlying set by insertions and deletions of elements. Also, the user can optimally verify in constant time the authenticity of the answer provided by a directory with a simple and practical algorithm. In particular, we show how to perform updates and queries in $O(n^{1/2})$ time while keeping the constant-time verification algorithm exactly the same as in previous inefficient schemes. In addition, at the expense of slightly increasing the conceptual complexity of the verification, we show that there is an accumulator-based approach to the authenticated dictionary problem that achieves $O(n^\epsilon)$-time performance for updates and queries, while keeping $O(1)$ verification time, where $\epsilon$ is any fixed constant such that $\epsilon > 0$. We have also implemented this scheme and we give empirical results that can be used to determine the best strategy for systems implementation with respect to resources that are available. This work has applications to certificate revocation in public key infrastructure and end-to-end integrity of data collections published by third parties on the Internet.

## 1  Introduction

Modern distributed transactions often operate in an asymmetric computational environment. Typically, client applications are deployed on small devices, such as laptop computers and palm devices, whereas the server side of these applications are often deployed on large-scale multiprocessors. Moreover, several client applications communicate with powerful server farms over wireless connections or slow modem-speed connections. Thus, distributed applications are facilitated by solutions that involve small amounts of computing and communication on the client side, without overly burdening the more-powerful server side of these same applications. The challenge we address in this paper is how to incorporate added levels of information assurance and security into such applications without significantly increasing the amount of computation and communication that is needed at the client (while at the same time keeping the computations on the servers reasonable).

A major aspect of our approach to this challenge is to replicate the computations of servers throughout mirror sites in the network, so as to reduce the network latency experienced by users in their client applications. This approach is used, for example, by Akamai Technologies to push images and other content to web servers that are close to client browsers. Thus, a user will in general be much closer to one of these mirror sites than to the source of the service, and will therefore experience a faster response time from a mirror than it would by communicating directly with the source. In addition, by off-loading user servicing from the source, this distributed scheme protects the source from denial-of-service attacks and allows for load balancing across the mirror sites, which further improves performance. Indeed, for the scope of this paper we are interested in supporting applications where clients can avoid online contact with the source.

[1]Dept. of Computer Science, University of California, Irvine. http://www.ics.uci.edu/~goodrich/. This work was supported in part by DARPA under grant F30602–00–2–0509, and by NSF under grants CCR-0311720, CCR-0312760, OCI-0724806, and IIS-0713046.

[2]Dept. of Computer Science, Brown University, Providence, Rhode Island. http://www.cs.brown.edu/~rt/. This work was supported in part by DARPA under grant F30602–00–2–0509, and by NSF under grants IIS–0713403 and OCI–0724806.

An information security problem arising in the replication of data to mirror sites is the authentication of the information provided by the sites. Indeed, there are applications where the user may require that data coming from a mirror site be cryptographically validated as being as genuine as they would be had the response come directly from the source. For example, a financial speculator that receives NASDAQ stock quotes from the *Yahoo! Finance* Web site would be well advised to get a proof of the authenticity of the data before making a large trade.

For all data replication applications, and particularly for e-commerce applications in wireless computing, we desire solutions that involve short responses from a mirror site that can be quickly verified with low computational overhead.

## 1.1 Problem Definition

More formally, the problem we address involves three groups of related parties: trusted information sources, untrusted directories, and users. An information *source* defines a finite set $S$ of elements that evolves over time through insertions and deletions of items. *Directories* maintain copies of the set $S$. Each directory storing $S$ receives time-stamped updates from the source for $S$ together with *update authentication information*, such as signed statements about the update and the current elements of the set. A *user* performs membership queries on the set $S$ of the type "is element $e$ in set $S$?" but instead of contacting the source for $S$ directly, it queries a directory for $S$ instead. The contacted directory provides the user with a response to the query together with *query authentication information*, which yields a proof of the answer assembled by combining statements signed by the source. The user then verifies the proof by relying solely on its trust in the source and the availability of public information about the source that allows to check the source's signature. The data structures used by the source directory to maintain set $S$, together with the protocol for queries and updates is called an *authenticated dictionary* [39]. Figure 1 shows a schematic view of an authenticated dictionary.

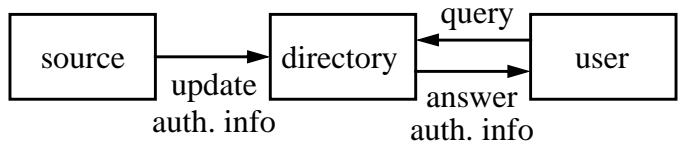

Figure 1: Authenticated dictionary.

The design of an authenticated dictionary should address several goals. These goals include low computational cost, so that the computations performed internally by each entity (source, directory, and user) should be simple and fast, and low communication overhead, so that bandwidth utilization is minimized. Since these goals are particularly important for the user, we say that an authenticated dictionary is *size-oblivious* if the query authentication information size and the verification time do not depend on the number of items in the dictionary. Size-oblivious solutions to the authenticated dictionary problem are ideally suited for wireless e-commerce applications, where user devices have low computational power and low bandwidth. In addition, size-oblivious solutions add an extra level of security, since the size of the dictionary is never revealed to users.

## 1.2 Applications

Authenticated dictionaries have a number of applications. One such application is in third-party data publication on the Internet [19], where third parties publish critical information, catalog entries, and design specifications for content providers who wish to outsource the business of publishing this information and processing transactions involving it.

In this case, the players in our framework are as follows: the source is a trusted organization (e.g., a stock exchange) that produces and maintains integrity-critical content (e.g., stock prices) and allows third parties (e.g., Web portals), to publish this content on the Internet so that it is widely disseminated. The publishers store copies of the content produced by the source and process queries on such content made by the users. In addition to returning the result of a query, a publisher also returns a proof of authenticity of the result, thus providing a validation service. Publishers also perform content updates originating from the source. Even so, the publishers provide this added value and are able to charge for it without the added cost of deploying all the mirror sites in high-security firewall-protected environments. Indeed, the publishers are not assumed to be trustworthy, for a given publisher may be processing updates from the source incorrectly or it may be the victim of a system break-in.

Another application of the authenticated dictionary is in certificate revocation [1, 13, 22, 27, 29, 38, 39], where the source is a *certification authority* (CA) that digitally signs certificates binding entities to their public keys, thus guaranteeing their validity. These certificates are then used to authorize secure socket layer (SSL) connections to e-stores and business-to-business exchanges. Nevertheless, certificates are sometimes revoked (e.g., if if a private key is lost or compromised, or if someone loses their authority to use a particular private key). Thus, the user of a certificate must be able to verify that a given certificate has not been revoked.

To facilitate such queries, the set of revoked certificates is distributed to *certificate revocation directories*, which process revocation status queries on behalf of users. The results of such queries need to be trustworthy, for they often form the basis for electronic commerce transactions.

Finally, we highlight how authenticated dictionaries could be used in military and research applications, for they could be used for the authenticated querying of information repositories, such as coalition documents, mission logs, genomic databases [28], and astrophysical databases (like the object catalog of the Sloan Digital Sky Survey [12, 32, 33]). Given the significant defense and scientific benefits that can result from such querying, users need to be certain that the results of their queries are accurate and current.

## 1.3 Previous and Related Work

Authenticated dictionaries are related to research in distributed computing (e.g., data replication in a network [7, 31]), data structure design (e.g., program checking [8, 10, 11, 43] and memory checking [9, 20]), and cryptography (e.g., incremental cryptography [4, 5, 20, 21]).

Previous additional work on authenticated dictionaries has been conducted primarily in the context of certificate revocation. The traditional method for certificate revocation (e.g., see [29]) is for the CA (source) to sign a statement consisting of a timestamp plus a hash of the set of all revoked certificates, called *certificate revocation list* (CRL), and periodically send the signed CRL to the directories. This approach is secure, but it is inefficient, for it requires the transmission of the entire set of revoked certificates for both source-to-directory and directory-to-user communication. Thus, this solution is clearly not size-oblivious, and even more recent modifications of this solution, which are based on delta-CRLs [17], are not size-oblivious.

Micali [38] proposes an alternate approach, where the source periodically sends to each directory the list of all issued certificates, each tagged with the signed timestamped value of a one-way hash function (e.g., see [42]) that indicates if this certificate has been revoked or not. This approach allows the system to reduce the size of the query authentication information to $O(1)$ words: namely just a certificate identifier and a hash value indicating its status. Unfortunately, this scheme requires the size of the update authentication information to increase to $\Theta(N)$, where $N$ is the number of all nonexpired certificates issued by the certifying authority, which is typically much larger than the number $n$ of revoked certificates. It is size-oblivious for immediate queries, but cannot be used for time stamping for archiving purposes, since no digest of the collection is ever made.

The *hash tree* scheme introduced by Merkle [36, 37] can be used to implement a static authenticated dictionary, which supports the initial construction of the data structure followed by query operations, but not update operations. A hash tree $T$ for a set $S$ stores the elements of $S$ at the leaves of $T$ and a hash value $h(v)$ at each node $v$, which combines the hash of its children. The authenticated dictionary for $S$ consists of the hash tree $T$ plus the signature of a statement consisting of a timestamp and the value $h(r)$ stored of the root $r$ of $T$. An element $e$ is proven to belong to $S$ by reporting the values stored at the nodes on the path in $T$ from the node storing $e$ to the root, together with the values of all nodes that have siblings on this path. This solution is not size-oblivious, since the length of this path depends on $n$. Kocher [30] also advocates a static hash tree approach for realizing an authenticated dictionary, but simplifies somewhat the processing done by the user to validate that an item is not in the set $S$, by storing intervals instead of individual elements. Other certificate revocation schemes, based on variations of cryptographic hashing, have been recently proposed in [13, 22], but like the static hash tree, these schemes have logarithmic verification time.

Using techniques from incremental cryptography, Naor and Nissim [39] dynamize hash trees to support the insertion and deletion of elements. In their scheme, the source and the directory maintain identically-implemented 2-3 trees. Each leaf of such a 2-3 tree $T$ stores an element of set $S$, and each internal node stores a one-way hash of its children's values. Hence, the source-to-directory communication is reduced to $O(1)$ items, but the directory-to-user communication remains at $O(\log n)$, where $n$ is the size of set $S$. Hence, their solution is also not size-oblivious.

Goodrich and Tamassia [24] have devised a data structure for an authenticated dictionary based on skip lists [40]. This data structure matches the asymptotic performance of the Naor-Nissim approach [39], while simplifying the details of an actual implementation of a dynamic authenticated dictionary.

Goodrich, Tamassia and Schwerin [26] present the software architecture and implementation of an authenticated dictionary based on the above approach, and Anagnostopoulos, Goodrich and Tamassia [2] introduce the notion of *persistent authenticated dictionaries*, where user can issue historical queries of the type, "was element $e$ in set $S$ at time $t$?"

Martel *et al.* [34] introduce a general approach for the design of authenticated data structures. They consider the class of data structures such that the ($i$) links of the structure form a directed acyclic graph $G$ of bounded degree and with a single source node; and ($ii$) queries on the data structure correspond to a traversal of a subdigraph of $G$ starting at the source. They show that such data structures can be authenticated by means of hashing scheme that digests the entire digraph $G$ into a hash value at its source. With this scheme, the sizes of the answer authentication information and the verification time are proportional to the size of the subdigraph traversed. Thus, their approach is not size-oblivious. Along these same lines, Cohen, Goodrich, Tamassia and Triandopoulos [16] show how to efficiently authenticate data structures for fundamental problems on networks, such as path and connectivity queries, and on geometric objects, such as intersection and containment queries. Their algorithms are more general than those of Martel *et al.*, but they still are not size-oblivious.

Independent of the preliminary announcement of the current paper [25], Camenisch and Lysyanskaya [14] independently investigate dynamic accumulators. They give a zero-knowledge protocol and a proof that a committed value is in the accumulator with respect to the Pedersen commitment scheme. They also present applications to revocation for group signature, identity escrow schemes and anonymous credentials systems. They do not achieve the kinds of performance tradeoffs we achieve in this paper.

## 1.4 Our Results

In this paper we present a number of size-oblivious solutions to the authenticated dictionary problem. The general approach we follow here is to abandon the approach of the previous methods cited above that are based on applying one-way hash functions to nodes in a data structure. Instead, we make use of RSA one-way accumulators, as advocated by Benaloh and de Mare [6]. Such an approach is immediately size-oblivious, but there is an additional challenge that has to be overcome to make this approach practical. The computations needed at the source and/or directories in a straightforward implementation of the Benaloh-de Mare scheme are inefficient. Our main contribution, therefore, is a mechanism to make the computations at the source and mirrors efficient.

The rest of this paper is organized as follows. In Section 2 we review the RSA accumulator [6] and other concepts used in our approach. We also present some basic tools that are used in the rest of the paper, including a description of a straightforward application of the RSA accumulator to the authenticated dictionary problem. We describe an improvement of this scheme that gives constant query and verification times but linear update time in Section 3. This improvement, called *precomputed accumulations*, consists of an efficient precomputation by the source of auxiliary data used by the directories to speed-up query processing. In Section 4, we present our complete solution, while preserving constant verification time by the user. For example, we can balance the two times and achieve $O(\sqrt{n})$ query and update time and $O(1)$ verification time, where $n$ is the current number of elements. An alternative solution is presented in Section 5, where we present a *parameterized accumulations* scheme. This scheme, suitable for large data sets, achieves $O(n^\epsilon)$ query and update time and $O(1)$ verification time, where $\epsilon$ is any fixed constant such that $\epsilon > 0$. Section 6 discusses the security of our scheme. In Section 7 we present the performance of our implementation of the scheme. Finally, concluding remarks are given in Section 8. Throughout this paper, we denote with $n$ the current number of elements of the set $S$ stored in the authenticated dictionary.

## 2 Preliminaries

In this section, we discuss some preliminary concepts used in our constructions.

### 2.1 The Security Model

As discussed above, in the *authenticated dictionary* problem, there are three parties, source, directory, and user, and a set $S$ of elements that are of interest to these three parties. Specifically, the roles of the three parties are as follows:

- The *source*: this is a trusted entity who is the author and authenticator for set $S$. The source maintains the set $S$ and vouches for its accuracy and content. The set $S$ is allowed to change over time, with insertions adding elements to $S$ and deletions removing items from $S$. At regular time intervals, the source communicates to the directory these changes together with a signed statement $(s, t)$, which we call *basis*, consisting of a timestamp $t$ and a value $s$ associated with the contents of set $S$ at time $t$. We denote with $\Delta$ the difference between two consecutive timestamps. Parameter $\Delta$ and the public key of the source are known to all the parties.

- The *directory*: this is an untrusted entity that periodically receives from the source the updates on set $S$ and the new signed basis for $S$. The directory also answers membership queries issued by a user. A membership query consists of asking whether a query item $e$ is in the most recent version of set $S$. The directory returns to the user a response $R_e$ ("$e \in S$" or "$e \notin S$") together with a proof of the response, consisting of a verification statement $C_e$ and the latest signed basis $(s, t)$ from the source. The directory must always answer each query in this way, but it can attempt to forge false answers and verification certificates.

- The *user*: this is an entity that issues a membership query for an element $e$ to the directory. After receiving the response $R_e$, statement $C_e$ and the signed basis $(s, t)$, the user verifies the signature of the basis and then runs a verification algorithm that takes as input $e$, $R_e$, $C_e$ and $s$ and outputs *true* if and only if $R_e$ is the correct answer to the query about element $e$ for the dictionary $S$ at time $t$. Finally, the user determines the timeliness of the response by checking that the current time is between the returned timestamp $t$ and the next timestamp $t + \Delta$.

In this paper, we are particularly interested in schemes for the authenticated dictionary problem such that the size of the verification statement and the running time of the verification algorithm are constant.

### 2.2 Reducing Dynamic Membership Determination to Dynamic Membership Verification

The authenticated dictionary problem requires the validation of two-sided answers, that is, whether "$e \in S$" or "$e \notin S$". As observed by Kocher [30], any authentication scheme that can perform secure membership verification, that is, one that can provide verification of "$e \in S$" statements, under element insertions and deletions, can be extended to provide verification of set membership determination.

Let $h$ be a collision-resistant cryptographic hash function mapping the universe of possible elements for set $S$ to $K$-bit integers. We show that can derive a membership determination scheme for set $S$ from a membership verification scheme for $2K$-bit integers. Let $X = (x_1, \cdots, x_n)$ be the sorted sequence of hash values of the elements of $S$ using hash function $h$. We define

$$Y = \{x_0 || x_1,\ x_1 || x_2,\ \cdots\ x_{n-1} || x_n,\ x_n || x_{n+1}\},$$

where $||$ denotes concatenation, $x_0 = 0$, and $x_{n+1} = 2^K - 1$. Each element $y_i = x_i || x_{i+1}$ of $Y$ represents the interval of hash values $[x_i, x_{i+1}]$. An insertion in set $S$ corresponds to two insertions and one deletion in set $Y$. Similarly, a deletion from set $S$ corresponds to one insertion and two deletions in set $Y$.

Suppose we have a membership verification scheme for $Y$. We can build a membership determination scheme for $S$ as follows. To prove $e \in S$, we return the proof of either $x_i || x_{i+1} \in Y$ or $x_{i-1} || x_i \in Y$, where $x_i = h(e)$. To prove that $e \notin S$, we return the proof that $x_i || x_{1+1} \in Y$, where $x_i < h(e) < x_{i+1}$. Therefore, for the remainder of this paper, we concentrate on set membership verification since we can reduce set membership verification to set membership determination.

## 2.3 One-Way Accumulators

An important tool for our scheme is that of one-way *RSA accumulator* functions [3, 6, 14, 23, 41]. Such a function allows a source to digitally sign a collection of objects as opposed to a single object.

The use of one-way RSA accumulators originates with Benaloh and de Mare [6]. They show how to utilize an RSA one-way accumulator, which is also known as an exponential accumulator, to summarize a collection of data so that user verification responses have constant-size. Refinements of the RSA accumulator used in our construction are given by Baric and Pfitzmann [3], Gennaro, Halevi and Rabin [23], and Sander, Ta-Shma and Yung [41].

As we show in the rest of this section, the RSA accumulator can be used to implement a static authenticated dictionary, where the set of elements is fixed. However, in a dynamic setting where items are inserted and deleted, the standard way of utilizing the RSA accumulator is inefficient. Several other researchers have also noted the inefficiency of this implementation in a dynamic setting (e.g., see [42]). Indeed, our solution can be viewed as refuting this previous intuition to show that a more sophisticated utilization of the RSA accumulator can be made to be efficient even in a dynamic setting.

The most common form of one-way accumulator is defined by starting with a "seed" value $y_0$, which signifies the empty set, and then defining the accumulation value incrementally from $y_0$ for a set of values $X = \{x_1, \cdots, x_n\}$, so that $y_i = f(y_{i-1}, x_i)$, where $f$ is a one-way function whose final value does not depend on the order of the $x_i$'s (e.g., see [6]). In addition, one desires that $y_i$ not be much larger to represent than $y_{i-1}$, so that the final accumulation value, $y_n$, is not too large. Because of the properties of function $f$, a source can digitally sign the value of $y_n$ so as to enable a third party to produce a short proof for any element $x_i$ belonging to $X$—namely, swap $x_i$ with $x_n$ and recompute $y_{n-1}$ from scratch—the pair $(x_i, y_{n-1})$ is a cryptographically-secure assertion for the membership of $x_i$ in set $X$.

A well-known example of a one-way accumulator function is the *RSA accumulator*,

$$f(y, x) = y^x \bmod N, \qquad (1)$$

for suitably-chosen values of the seed $y_0$ and modulus $N$ [6]. In particular, choosing $N = PQ$ with $P$ and $Q$ being two strong primes [35] makes the RSA accumulator function as difficult to invert as RSA cryptography [6].

The difficulty in using the RSA accumulator function in the context of authenticated dictionaries is that it is not associative; hence, any updates to set $X$ require significant recomputations.

## 2.4 Euler's Theorem

There is an important technicality involved with use of the RSA accumulator function, namely in the choice of the seed $a = y_0$. In particular, we should choose $a$ relatively prime with $P$ and $Q$. This choice is dictated by Euler's Theorem, which states

**Theorem 2.1** (Euler's Theorem). $a^{\phi(N)} \bmod N = 1$, *if $a > 1$ and $N > 1$ are relatively prime.*

In our use of the RSA accumulator function, the following well-known corollary to Euler's Theorem will prove useful.

**Corollary 2.2.** *If $a > 1$ and $N > 1$ are relatively prime, then $a^x \bmod N = a^{x \bmod \phi(N)} \bmod N$, for all $x \geq 0$.*

One implication of this corollary to the authenticated dictionary problem is that the source should never reveal the values of the prime numbers $P$ and $Q$. Such a revelation would allow a directory to compute $\phi(N) = (P-1)(Q-1)$, which in turn could result in a false validation at a compromised directory. So, our approach takes care to keep the values of $P$ and $Q$ only at the source.

## 2.5 Two-Universal Hash Functions

As in previous approaches [23, 41], we use the RSA accumulator in conjunction with two-universal hash functions. Such functions were first introduced by Carter and Wegman [15].

A family $H = \{h : A \to B\}$ of functions is *two-universal* if, for all $a_1, a_2 \in A$, $a_1 \neq a_2$ and for a randomly chosen function $h$ from $H$,

$$\Pr_{h \in H}\{h(a_1) = h(a_2)\} \leq \frac{1}{|B|}.$$

In our scheme, the set $A$ consists of $3k$-bit vectors and the set $B$ consists of $k$-bit vectors, and we are interested in finding random elements in the preimage of a two-universal function mapping $A$ to $B$. We can use the two-universal function $h(x) = Ux$, where $U$ is a $k \times 3k$ binary matrix. To get a representation of all the solutions for $h^{-1}(e)$, we need to solve a linear system. Once this is done, picking a random solution can be done by multiplying a bit matrix by a random bit vector, and takes $O(k^2)$ bit operations.

## 2.6 Choosing a Suitable Prime

We are interested in obtaining a prime solution of the linear system that represents a two-universal hash function. The following lemma, of Gennaro *et al.* [23], is useful in this context:

**Lemma 2.3** ([23]). *Let $H$ be a two-universal family from $\{0,1\}^{3k}$ to $\{0,1\}^k$. Then, for all but a $2^{-k}$ fraction of the functions $h \in H$, for every $e \in \{0,1\}^k$ a fraction of at least $\frac{1}{ck}$ of the elements in $f^{-1}(e)$ are primes, for some small constant c.*

For reasons that will become clear in the security proof given in Section 6, our scheme requires that a prime inverse be greater then $\sqrt{2^{3k}}$. Also, since the domain of $H$ is $\{0,1\}^{3k}$, this prime is less than $2^{3k}$. So, by the results of prime number theory, the density of big prime numbers that are less than $2^k$ is about $\frac{1}{2k}$ for all but a $2^{-\Omega(k)}$ fraction of functions in family $H$. The expected number of steps to find a suitable prime is $O(k)$. In order to find a suitable prime with high probability $1 - 2^{-\Omega(k)}$ we need to sample $O(k^2)$ times.

Recall from Section 2.5 that picking a random solution takes $O(k^2)$ bit operations. Thus, the total running time of finding a suitable prime is equal to running $O(k^2)$ primality tests.

One needs to be careful about choice of primality test because it could happen that the cost of prime generation and verification dominates the cost of signing. One could use the Miller-Rabin test, for example. To reduce the probability of mistaking a composite number for a prime one could perform a number of additional Miller-Rabin tests. Performing these tests could be costly. Fortunately, Cramer and Shoup [18] give a fast primality testing algorithm that can be used here. It does additional tests between runs of Miller-Rabin algorithm that reduce the primality checking time. They also state that empirical runs of the algorithm indicate running times that are suitable for signing schemes.

## 2.7 The Strong RSA Assumption

The proof of security of our scheme uses the *strong RSA assumption*, as defined by Baric and Pfitzmann [3]. Given $N$ and $x \in Z_N^*$, the strong RSA problem consists of finding integers $f$, with $2 \le f < N$, and $a$, such that we have $a^f = x$. The difference between this problem and the standard RSA problem is that the adversary is given the freedom to choose not only the base $a$ but also the exponent $f$.

> *Strong RSA Assumption:* There exists a probabilistic algorithm $B$ that on input $1^r$ outputs an RSA modulus $N$ such that, for all probabilistic polynomial-time algorithms $D$, all $c > 0$, and all sufficiently large $r$, the probability that algorithm $D$ on a random input $x \in Z_N$ outputs $a$ and $f \ge 2$ such that $a^f = x \bmod N$ is no more than $r^{-c}$.

In other words, given $N$ and a randomly chosen element $x$, it is infeasible to find $a$ and $f$ such that $a^f = x \bmod N$.

## 2.8 A Straightforward Accumulator-Based Scheme

Let $S = \{e_1, e_2, \ldots, e_n\}$ be the set of elements stored at the source. Each element $e$ is represented by $k$ bits. The source chooses strong primes [35] $P$ and $Q$ that are suitably large, e.g., $P, Q > 2^{\frac{3}{2}k}$. It then chooses a suitably-large base $a$ that is relatively prime to $N = PQ$. Note that $N$ is at least $2^{3k}$. It also chooses a random hash function $h$ from a two-universal family (as discussed in Section 2.5). The source broadcasts once $a$, $N$ and $h$ to the directories and users, but keeps $P$ and $Q$ secret. At periodic time intervals, for each element $e_i$ of $S$, the source computes the *representative* of $e_i$, denoted $x_i$, where $h(x_i) = e_i$ and $x_i$ is a prime chosen as described in Section 2.6. The source then combines the representatives of the elements by computing the RSA accumulation

$$A \leftarrow a^{x_1 x_2 \cdots x_n} \bmod N$$

and broadcasts to the directories a signed message $(A, t)$, where $t$ is the current timestamp.

### 2.8.1 Query

When asking for a proof of membership in $S$ of an element $e_i$, the user submits $e_i$ to a directory. To prove that a query element $e_i$ is in $S$, a directory computes the value

$$A_i \leftarrow a^{x_1 x_2 \cdots x_{i-1} x_{i+1} \cdots x_n} \bmod N. \qquad (2)$$

That is, $A_i$ is the accumulation of all the representatives of the elements of $S$ besides $x_i$ and is said to be the *witness* of $e_i$. After computing $A_i$, the directory returns to the user the representative $x_i$, the witness $A_i$ and the pair $(A, t)$, signed by the source. Note that this query authentication information has constant size; hence, this scheme is size-oblivious. However, computing witness $A_i$ is no trivial task for the directory, for it must perform $n - 1$ exponentiations to answer a query. Making the simplifying assumption that the number of bits needed to represent $N$ is independent of $n$, the computation performed to answer a single query takes $O(n)$ exponentiations.

### 2.8.2 Verification

The user checks that timestamp $t$ is current and that $(A, t)$ is indeed signed by the source. It then checks that $x_i$ is a valid representative of $e_i$, i.e., $h(x_i) = e_i$. Finally, it computes $A' \leftarrow A_i^{x_i} \bmod N$ and compares it to $A$. If $A' = A$, then the user is reassured of the validity of the answer because of the strong RSA assumption. The verification needs only $O(1)$ exponentiations.

### 2.8.3 Updates

For updates, the above simple approach has an asymmetric performance (for unrestricted values of accumulated elements), with insertions being much easier than deletions. To insert a new element $e_{n+1}$ into the set $S$, the source simply recomputes the accumulation $A$ as follows

$$A \leftarrow A^{x_{n+1}} \bmod N$$

where $x_{n+1}$ is the (prime) representative of $e_{n+1}$. The computation of $x_{n+1}$ can be done in time that is independent of $n$ (see Section 2.6), i.e., with one exponentiation. An updated signed pair $(A, t)$ is then sent to the directories in the next time interval. Thus, an insertion takes $O(1)$ time, counting exponentiations and other modular arithmetic operations as constant-time operations. The deletion of an element $e_i \in S$, on the other hand, will in general require the source to recompute the new value $A$ by performing $n - 1$ exponentiations. That is, a deletion takes $O(n)$ time.

The performance of the above straightforward use of the RSA accumulator is summarized in Table 1.

| space | insertion time | deletion time | update info | query time | query info | verify time |
|-------|----------------|---------------|-------------|------------|------------|-------------|
| $O(n)$ | $O(1)$ | $O(n)$ | $O(1)$ | $O(n)$ | $O(1)$ | $O(1)$ |

Table 1: Straightforward implementation of an authenticated dictionary using an RSA accumulator. Each of these running times count modular exponentiations and other arithmetic operations as constant-time operations; hence, they can also be viewed in terms of alternative bounds by using the above bounds as characterizing the number of such modular operations.

Of course, if a representative $x_i$ is relatively prime with $P - 1$ and $Q - 1$, the source can delete $e_i$ by computing $x \leftarrow x_i^{-1} \bmod \phi(N)$ (via the extended Euclidean algorithm) and then updating $A \leftarrow A^x \bmod N$. But we cannot guarantee that $x_i$ has an inverse in $Z_{\phi(N)}$ if it is an accumulation of a group of elements; hence, we do not advocate using this approach for deletions. Indeed, we will not assume the existence of multiplicative inverses in $Z_{\phi(N)}$ for any of our solutions. Thus, we are stuck with linear deletion time at the source and linear query time at a directory when making this straightforward application of RSA accumulators to the authenticated dictionary problem.

The above linear query time is generally considered too slow to be efficient for processing large numbers of queries. We describe in the next section an alternative approach that can answer queries much faster.

## 3 Precomputed Accumulations

We present a first improvement that allows for fast query processing. We require the directory to store a precomputed witness $A_i$ for each element $e_i$ of $S$, as defined in Eq. 2. Thus, to answer a query, a directory looks up the $A_i$ value, rather than computing it from scratch, and then completes the transaction as described in the previous section. Thanks to the precomputation of the witnesses at the source, a directory can process any query in $O(1)$ time (with no exponentiations) while the verification computation for a user remains unchanged.

Unfortunately, a standard way of implementing this approach is inefficient for processing updates. In particular, a directory now takes $O(n)$ exponentiations to process a single insertion, since it needs to update all the existing witnesses and compute a new witness from scratch, and $O(n \log n)$ exponentiations to process a single deletion, for after a deletion the directory must recompute all the witnesses, which can be done using the algorithm given in [41]. Thus, at first glance, this precomputed accumulations approach appears to be quite inefficient when updates to the set $S$ are required.

We can process updates with fewer than $O(n \log n)$ exponentiations, however, by enlisting the help of the source. Our method in fact can be implemented with $O(n)$ exponentiations by a simple two-phase approach. The details for the two phases follows.

## 3.1 First Phase

Let $S$ be the set of $n$ items stored at the source after performing all the insertions and deletions required in the previous time interval. We build a complete binary tree $T$ "on top" of the representative values of the elements of $S$, so that each leaf of $T$ is associated with the representative $x_i$ of an element $e_i$ of $S$. In the first phase, we perform a post-order traversal of $T$, so that each node $v$ in $T$ is visited only after its children are visited. The main computation performed during the visit of a node $v$ is to compute a value $x(v)$. If $v$ is a leaf of $T$, storing some representative $x_i$, then we compute

$$x(v) \leftarrow x_i \bmod \phi(N).$$

If $v$ is an internal node of $T$ with children $u$ and $w$ (we can assume $T$ is proper, so that each internal node has two children), then we compute

$$x(v) \leftarrow x(u)x(w) \bmod \phi(N).$$

When we have computed $x(r)$, where $r$ denotes the root of $T$, then we are done with this first phase. Since a post-order traversal takes $O(n)$ time, and each visit computation in our traversals takes $O(1)$ time, this entire first phase runs in $O(n)$ time. We again make the simplifying assumption that the number of bits needed to represent $N$ is independent of $n$.

## 3.2 Second Phase

In the second phase, we perform a pre-order traversal of $T$, where the visit of a node $v$ involves the computation of a value $A(v)$. The value $A(v)$ for a node $v$ is defined to be the accumulation of all values stored at nodes that are *not* descendents of $v$ (including $v$ itself if $v$ is a leaf). Thus, if $v$ is a leaf associated with the representative value $x_i$ of some element of $S$, then $A(v) = A_i$. Recall that in a pre-order traversal, we perform the visit action on each node $v$ before we perform the respective visit actions for $v$'s children. For the root, $r$, of $T$, we define $A(r) = a$. For any non-root node $v$, let $z$ denote $v$'s parent and let $w$ denote $v$'s sibling (and note that since $T$ is proper, every node but the root has a sibling). Given $A(z)$ and $x(w)$, we can compute the value $A(v)$ for $v$ as follows:

$$A(v) \leftarrow A(z)^{x(w)} \bmod N.$$

By Corollary 2.2, we can inductively prove that each $A(v)$ equals the accumulation of all the values stored at non-descendents of $v$. Since a pre-order traversal of $T$ takes $O(n)$ time, and each visit action can be performed with $O(1)$ exponentiations, we can compute all the $A_i$ witnesses with $O(n)$ exponentiations. Note that implementing this algorithm requires knowledge of the value $\phi(N)$, which presumably only the source knows. Thus, this computation can only be performed at the source, who must transmit the updated $A_i$ values to the directory.

The performance of the precomputed accumulation scheme is summarized in Table 2.

In the next section, we show how to combine this approach with the straightforward approach of Section 2.8 to design a scheme that is efficient for both updates and queries.

| space | insertion time | deletion time | update info | query time | query info | verify time |
|---|---|---|---|---|---|---|
| $O(n)$ | $O(n)$ | $O(n)$ | $O(n)$ | $O(1)$ | $O(1)$ | $O(1)$ |

Table 2: Precomputed accumulation scheme for implementing an authenticated dictionary with an RSA accumulator. Each of these running times count modular exponentiations and other arithmetic operations as constant-time operations; hence, they can also be viewed in terms of alternative bounds by using the above bounds as characterizing the number of such modular operations.

## 4  Parameterized Accumulations

Consider again the problem of designing an accumulator-based authenticated dictionary for a set $S = \{e_1, e_2, \ldots, e_n\}$. In this section, we show how to balance the processing between the source and the directory, depending on their relative computational power. The main idea is to choose an integer parameter $1 \leq p \leq n$ and partition the set $S$ into $p$ groups of roughly $n/p$ elements each, performing the straightforward approach inside each group and the precomputed accumulations approach among the groups (see Figure 2). The details are as follows.

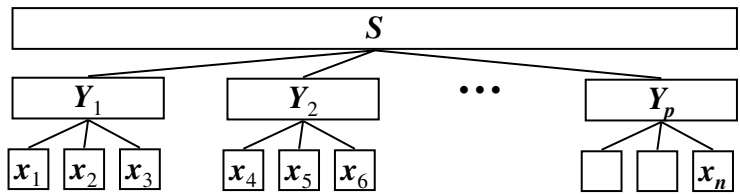

Figure 2: Parameterized accumulations scheme.

### 4.1  Subdividing the Dictionary

Divide the set $S$ into $p$ groups, $Y_1, Y_2, \ldots, Y_p$, of roughly $n/p$ elements each, balancing the size of the groups as much as possible. For group $Y_j$, let $y_j$ denote the product of the representatives of the elements in $Y_j$ modulo $\phi(N)$. Define $B_j$ as

$$B_j = a^{y_1 y_2 \cdots y_{j-1} y_{j+1} \cdots y_p} \bmod N.$$

That is, $B_j$ is the accumulation of representatives of all the elements that are not in the set $Y_j$. After any insertion or deletion in a set $Y_j$, the source can compute a new value $y_j$ with $O(n/p)$ exponentiations. (We show in Section 4.2 how with some effort this bound can be improved to $O(\log(n/p))$ exponentiations).) Moreover, since the source knows the value of $\phi(N)$, it can update all the $B_j$ values after such an update in $O(p)$ time. Thus, the source can process an update operation in $O(p + n/p)$ time, assuming that the update does not require redistributing elements among the groups and we are counting modular exponentiations as constant-time operations.

Maintaining the size of each set $Y_j$ is not a major overhead. We need only keep the invariant that each $Y_j$ has at least $\lceil n/p \rceil / 2$ elements at most $2\lceil n/p \rceil$ elements. If a $Y_j$ set becomes too small, then we either merge it with one of its adjacent sets $Y_{j-1}$ or $Y_{j+1}$, or (if merging $Y_j$ with such a sets would cause an overflow) we "borrow" some of the elements from an adjacent set to bring the size of $Y_j$ to at least $3\lceil n/p \rceil / 4$. Likewise, if a $Y_j$ set grows too large, then we simply split it in two. These simple adjustments take $O(n/p)$ time, and will maintain the invariant that each $Y_j$ is of size $\Theta(n/p)$. Of course, this assumes that the value of $n$ does not change significantly as we insert and remove elements. But even this condition is easily handled. Specifically, we can maintain the sizes of the $Y_j$'s in a priority queue that keeps track of the smallest and largest $Y_j$ sets. Whenever we increase $n$ by an insertion, we can check the priority queue to see if the smallest set now must do some merging or borrowing to keep from growing too small. Likewise, whenever we decrease $n$ by a deletion, we can check the priority queue to see if the largest set now must split. An inductive argument shows that this approach keeps the size of the groups to be $\Theta(n/p)$.

Keeping the $Y_j$'s to have exactly size $\Theta(n/p)$ is admittedly an extra overhead. In practice, however, all this overhead can probably be ignored, as it is likely that the $Y_j$'s will grow and shrink at more or less the same rate. Indeed, even if the updates are non-uniform, we can afford to completely redistribute the elements in all the $Y_j$'s as often as every $O(\min\{p, n/p\})$ updates, amortizing the $O(n)$ cost for this redistribution to the previous set of updates that occurred since the last redistribution.

Turning to the task at a directory, then, we recall that a directory receives all $p$ of the $B_j$ values after an update occurs. Thus, a directory can perform its part of an update computation in $O(p)$ time. It validates that some $e_i$ is in $e$ by first determining the group $Y_j$ containing $e_i$, which can be done by table look-up. Then, it computes $A_i$ as

$$A_i \leftarrow B_j^{\Pi_{e_m \in Y_j - \{e_i\}} x_m} \mod N,$$

where $x_m$ is the representative of $e_m$. Thus, a directory can answer a query with $O(n/p)$ exponentiations.

The performance of the parameterized accumulation algorithm is summarized in Table 3.

| space | insertion time | deletion time | update info | query time | query info | verify time |
|-------|----------------|---------------|-------------|------------|------------|-------------|
| $O(n)$ | $O(p + n/p)$ | $O(p + n/p)$ | $O(p)$ | $O(n/p)$ | $O(1)$ | $O(1)$ |

Table 3: Parameterized accumulations scheme for implementing an authenticated dictionary using an RSA accumulator. We denote with $p$ an integer such that $1 \leq p \leq n$. Each of these running times count modular exponentiations and other arithmetic operations as constant-time operations; hence, they can also be viewed in terms of alternative bounds by using the above bounds as characterizing the number of such modular operations.

The parameter $p$ allows us to balance the work between the source and the directories, and also between updates and queries. For example, we can set $p = \lceil \sqrt{n} \rceil$, which gives $O(\sqrt{n})$ time for both queries and updates. Note that for reasonable values of $n$, say for $n$ between $10,000$ and $1,000,000$, $\sqrt{n}$ is between 100 and 1,000. In many cases, this is enough of a reduction to make the dynamic RSA accumulator practical for the source and directories, while still keeping the user computation to be one exponentiation and one signature verification. Indeed, these user computations are simple enough to even be embedded in a smart card, a PDA, or mobile phone.

## 4.2 Improving the Update Time for the Source

In this section, we show how the source can further improve the performance of an update operation in the parameterized scheme. Recall that in this scheme the set $S$ is partitioned into $p$ subsets, $Y_1, Y_2, \ldots, Y_p$, and the source maintains for each $Y_j$ a value $B_j$, on behalf of the directories, that is the accumulation of all the values not in $Y_j$. Also recall that, for each group $Y_j$, we let $y_j$ denote the product of the items in $Y_j$ modulo $\phi(N)$. In the algorithm described above, the source recomputes $y_j$ from scratch after any update occurs, which takes $O(n/p)$ exponentiations. We will now describe how this computation can be done with $O(\log(n/p))$ exponentiations.

The method is for the source to store the elements of each $Y_j$ in a balanced binary search tree. For each internal node $w$ in $T_j$, the source maintains the value $y(w)$, which is the product of the representatives of all the items stored at descendents of $w$, modulo $\phi(N)$. Thus, $y(r(T_j)) = y_j$, where $r(T_j)$ denotes the root of $T_j$. Any insertion or deletion will affect only $O(\log(n/p))$ nodes $w$ in $T_j$, for which we can recompute their $x(w)$ values in $O(\log(n/p))$ total time. Therefore, after any update, the source can recompute a $y_j$ value in $O(\log(n/p))$ time, assuming that the size of the $Y_j$'s does not violate the size invariant. Still, if the size of $Y_j$ after an update violates the size invariant, we can easily adjust it by performing appropriate splits and joins on the trees representing $Y_j$, $Y_{j-1}$, and/or $Y_{j+1}$. Moreover, we can rebuild the entire set of trees after every $O(n/p)$ updates, to keep the sizes of the $Y_j$ sets to be $O(n/p)$, with the cost for this periodic adjustment (which will probably not even be necessary in practice for most applications) being amortized over the previous updates. The performance of the resulting scheme is summarized in Table 4.

| space | insertion time | deletion time | update info | query time | query info | verify time |
|---|---|---|---|---|---|---|
| $O(n)$ | $O(p + \log(n/p))$ | $O(p + \log(n/p))$ | $O(p)$ | $O(n/p)$ | $O(1)$ | $O(1)$ |

Table 4: Enhanced parameterized scheme for implementing an authenticated dictionary using an RSA accumulator. We denote with $p$ an integer such that $1 \leq p \leq n$. Each of these running times count modular exponentiations and other arithmetic operations as constant-time operations; hence, they can also be viewed in terms of alternative bounds by using the above bounds as characterizing the number of such modular operations.

In this version of our scheme, we can achieve a complete tradeoff between the cost of updates at the source and queries at the directories. Tuning the parameter $p$ over time, therefore, could yield the optimal balance between the relative computational powers of the source and directories. It could also be used to balance between the number of queries and updates in the time intervals.

**Theorem 4.1.** *The parameterized accumulations scheme for implementing an authenticated dictionary over a set of size $n$ uses data structures with $O(n)$ space at the source and directories and has the following performance, for a given parameter $p$ such that $1 \leq p \leq n$:*

- *the insertion and deletion operations for the source each require $O(p + \log(n/p))$ exponentiations;*

- *the update authentication information has size $O(p)$;*

- *answering a query by a directory requires $O(n/p)$ exponentiations;*

- *the query authentication information has size $O(1)$; and*

- *the verification for a user requires only $O(1)$ exponentiations.*

Thus, for $p = \sqrt{n}$, one can balance insertion time, deletion time, update authentication information size, and query time to achieve an $O(\sqrt{n})$ bound, while keeping the query authentication information size and the verification time constant.

The parameterized accumulations scheme described in this section significantly improves the overhead at the source and directories for using an RSA accumulator to solve the authenticated dictionary problem. Moreover, this improvement was achieved without any modification to the client from the original straightforward application of the RSA accumulator described in Section 2.8.

In the next section, we show that if we are allowed to slightly modify the computation at the client, we can further improve performance at the source and directory while still implementing a size-oblivious scheme

## 5 Hierarchical Accumulations

In this section, we describe a hierarchical accumulation scheme for implementing an authenticated dictionary on a set $S$ with $n$ elements. In this scheme, the verification algorithm consists of performing a series of $c$ exponentiations, where $c$ is a fixed constant for the scheme (see Figure 3). Note that the approach of Section 4 assumed that $c = 1$(not counting the exponentiation needed to verify the source's digital signature of the pair $(A, t)$ if RSA signature scheme is used).

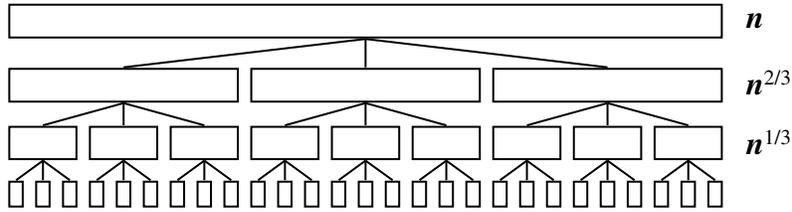

Figure 3: Hierarchical accumulations scheme with parameter $c = 2$.

Given a fixed constant $c$, we define $p = n^{1/(c+1)}$ and construct the following hierarchical partition of $S$:

- We begin by partitioning set $S$ into $p$ subsets of $n_1 = n^{c/(c+1)}$ elements each, called level-1 subsets.

- For $i = 1, \ldots, c-1$, we partition each level-$i$ subset into $p$ subsets of $n^{(c-i)/(c+1)}$ elements each, called level-$(i+1)$ subsets.

Also, we conventionally say that $S$ is the level-0 subset.

Next, we associate a value $\alpha(Y)$ to each subset $Y$ of the above partition, as follows:

- The value of a level-$c$ subset is the accumulation of the representatives of its elements.

- For $i = 0, \ldots, c-1$, the value of a level-$i$ subset is the accumulation of the representatives of the values of its level-$(i+1)$ subsets.

Finally, we store with each level-$i$ subset $Y$ a data structure based on the precomputed accumulations scheme of Section 3 that stores and validates membership in the set $S(Y)$ of the values of the level-$(i+1)$ subsets of $Y$.

Let $e$ be an element of $S$. To prove the containment of $e$ in $S$, the directory determines, for $i = 1, \ldots, c$, the level-$i$ subset $Y_i$ containing $e$ and returns the sequence of values $\alpha(Y_c), \alpha(Y_{c-1}), \ldots, \alpha(Y_0)$ plus witnesses for the following $c + 1$ memberships:

- $e \in Y_c$

- $\alpha(Y_i) \in S(Y_{i-1})$ for $i = c, \ldots, 1$

The user can verify each of the above memberships by means of an exponentiation. Thus, the verification time and query authentication information are proportional to $c$, i.e., they are $O(1)$.

The performance of the hierarchical accumulations scheme is summarized in Table 5.

| space | insertion time | deletion time | update info | query time | query info | verify time |
|---|---|---|---|---|---|---|
| $O(n)$ | $O(n^\epsilon)$ | $O(n^\epsilon)$ | $O(n^\epsilon)$ | $O(n^\epsilon)$ | $O(1)$ | $O(1)$ |

Table 5: Hierarchical accumulations scheme for implementing an authenticated dictionary with an RSA accumulator, where $1/(c+1) < \epsilon < 1$ is a fixed constant (there is a constant factor of $1/\epsilon$ "hiding" behind each of the big-ohs in this table). Each of these running times count modular exponentiations and other arithmetic operations as constant-time operations; hence, they can also be viewed in terms of alternative bounds by using the above bounds as characterizing the number of such modular operations.

The hierarchical accumulations scheme is likely to outperform in practice the parameterized accumulations scheme only for large-scale authenticated dictionaries (say, containing billions of entries), where the difference between $n^{1/(c+1)}$ and $n^{1/2}$ is significant and offsets the added complication of changing the client code and introducing the $(c+1)$-level accumulation hierarchy.

**Theorem 5.1.** *The hierarchical accumulations scheme for implementing an authenticated dictionary over a set of size $n$ uses uses data structures with $O(n)$ space at the source and directories and has the following performance, for a given constant $\epsilon$ such that $0 < \epsilon < 1$:*

- *the insertion and deletion operations for the source each can be done with $O(n^\epsilon))$ exponentiations;*
- *the update authentication information has size $O(n^\epsilon)$;*
- *a query at a directory can be answered with $O(n^\epsilon)$ exponentiations;*
- *the query authentication information has size $O(1)$; and*
- *the verification for a user requires only $O(1)$ exponentiations.*

We can extend the hierarchical accumulations scheme by using a more general hierarchical partitioning of the set $S$ while keeping constant the size of the query authentication information as well as the number of exponentiations for verification. The two extreme partitioning strategies are: (i) single-level partition in $O(1)$ groups of size $O(n)$, and (ii) $O(\log n)$-level partition where the size of each partition is $O(1)$ (this corresponds to a hierarchy that can be mapped into a bounded-degree tree). The insertion and deletion times and the update authentication information size are then proportional to $O(\sum_{i=1}^{c-1} g_i)$, where $g_i$ is the size of the partition at the $i$-th level, and can range from $O(1)$ to $O(n)$. At the same time, the query time is proportional to $O(c + g_{c-1})$, and can range from $O(n)$ to $O(1)$. The number of precomputed values that need to be stored affects the amount of space needed per element, which varies from $O(1)$ to $O(n)$. Thus, the space increase per element is $O(1)$, or to be more precise, at most two precomputed values can be stored per element in the dictionary (if the underlying hierarchy tree is binary).

# 6 Security

We now show that an adversarial directory cannot forge a proof of membership for an element that is not in $S$. Our proof follows a closely related constructions given in [14, 23, 41]. An important property of the scheme comes from representing the elements $e$ of set $S$ with prime numbers. If the accumulator scheme was used without this stage, the scheme would be insecure. An adversarial directory could forge the proof of membership for all the divisors of elements whose proofs it has seen.

**Theorem 6.1.** *In the dynamic accumulator schemes for authenticated dictionaries defined in the previous sections, under the strong RSA assumption, a directory whose resources are polynomially bounded can produce a proof of membership only for the elements that are in the dictionary.*

*Proof.* Our proof is based on related proofs given in [14, 23, 41]. Assume an adversarial directory $D$ has seen proofs of membership for all the elements $e_1, e_2, \ldots e_n$ of the dictionary $S$. The trusted source has computed representatives $x_1, x_2, \ldots, x_n$ as suitable primes defined in Section 2.6. The witnesses $A_1, A_2 \ldots, A_n$ have been computed as well, either solely by the trusted source, or by balancing the work between the trusted source and the directories. The trusted source has distributed a signed pair $(A, t)$. By the definition of the scheme in Section 2.6, for all $1 \leq i \leq n$, we have

- $x_i$ is the prime representative of $e_i \in S$, i.e., $h(x_i) = e_i$;
- $\sqrt{2^{3k}} < x_i < 2^{3k}$;
- $A_i^{x_i} \bmod N = A$.

We need to show that directory $D$ cannot prove the membership of an element $e_{n+1}$ that is not in the set $S$ already. The proof is by contradiction. Suppose that $D$ has has found a triplet $(e_{n+1}, x_{n+1}, A_{n+1})$ proving the membership of $e_{n+1}$. Then, the following must hold and can checked by the user (it is not necessary for $x_{n+1}$ to be a prime):

- $h(x_{n+1}) = e_{n+1}$;
- $\sqrt{2^{3k}} < x_{n+1} < 2^{3k}$;
- $A_{n+1}^{x_{n+1}} \bmod N = A$.

Let $d = \gcd(x_{n+1}, x_1 x_2 \ldots x_n)$. Thus, we have $\gcd(\frac{x_{n+1}}{d}, \frac{x_1 x_2 \ldots x_n}{d}) = 1$. Define $f = \frac{x_{n+1}}{d}$. There are integers $u, v$ such that $v \frac{x_1 x_2 \ldots x_n}{d} + uf = 1$ holds over integers. Directory $D$ can find $u$ and $v$ in polynomial time using the extended Euclidean algorithm. Set $s = A_{n+1}^v a^u$. We have

$$s^f = A_{n+1}^{vf} a^{uf} = A_{n+1}^{v \frac{x_{n+1}}{d}} a^{uf} = A^{\frac{v}{d}} a^{uf} = a^{v \frac{x_1 x_2 \ldots x_n}{d} + uf} = a.$$

Thus, directory $D$ can find in polynomial time a value $s$ that is an $f$-th root of $a$. By the strong RSA assumption (Section 2.7), it must be that $f = 1$. Hence, we have $x_{n+1} = d$ and it follows that $x_{n+1}$ divides $x_1 x_2 \ldots x_n$. But by our assumptions we have $x_{n+1} < 2^{3k}$ and $x_i > \sqrt{2^{3k}}$ for each $i$, which implies that $x_{n+1} = x_i$, for some $1 \leq i \leq n$. Thus, element $e_{n+1}$ is already in set $S$, which is a contradiction.

We conclude that the adversarial directory $D$ can find membership proofs only for those elements already in $S$. □

# 7 Experimental Results

In this section, we present a preliminary experimental study on the performance of the dynamic accumulator schemes for authenticated dictionaries described in this paper. The main results of this study are summarized in the the charts of Figures 4–5, where the $x$-axis represents the size of the dictionary (number of elements) and the $y$-axis represents the average time of the given operation in microseconds. We denote with $(f_1(n), f_2(n), ..., f_c(n))$ a generalized hierarchical partition scheme of the dictionary with $O(f_i(n))$ elements in the $i$-th level group (Section 5).

The dynamic accumulator scheme has been implemented in Java and the experiments have been conducted on an AMD Athlon XP 1700+ 1.47GHz, 512MB running Linux. The items stored in the dictionary and the query values are randomly generated 165-bit integers and the parameter $N$ of the RSA accumulator is a 200-bit integer. The variance between different runs of the query and deletion operations was found to consistently small so only a few runs were done for each dictionary size considered.

The main performance bottleneck of the scheme was found to be the computation of prime representatives for the elements. In our experiments, finding a prime representative of a 165-bit integer using the standard approach of Section 2.6 takes about 45 milliseconds and dominates the rest of the insertion time. The computation of prime representatives is a constant overhead that does not depend on the number of elements and has been omitted in the rest of the analysis.

Figure 4 illustrates two performance tradeoffs. Part (a) compares the performance of the two extreme naive approaches where either the source or the directory does essentially all the work. Since the source can use modular multiplication and the directory has to use modular exponentiation, it is more effective to shift as much as possible the insertion work to the source. Part (b) shows the benefits of partitioning, which allows to reduce the computation time at the source.

Experimental results on the hierarchical accumulations method (Section 5) are presented in Figure 5. These results show that one can tune the partitioning scheme according to the processing power available at the source and the directory. Thus, this experimental analysis shows that the 2-level $(n^{1/2}, n^{1/4})$ partitioning scheme is superior to the $(n^{2/3}, n^{1/3})$ partitioning scheme, and both are much better than unpartitioned schemes (which would have linear performance).

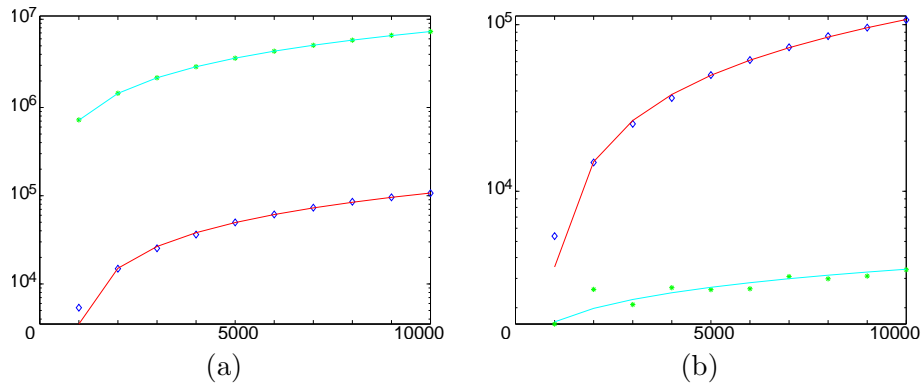

Figure 4: Performance tradeoffs for dynamic accumulators. The insertion time at the source excludes the computation of the prime representative. Note that we use a logarithmic scale for the $y$-axis. **(a)** Query time at the directory (stars) when the directory computes the witness from scratch for each query (using modular exponentiations) vs. insertion time at the source (diamonds) when the source precomputes all the exponents of the witnesses (using modular multiplications). **(b)** Insertion time at the source (diamonds) without partitioning, when all the $n$ witnesses's exponents are precomputed, vs. with partitioning (stars), when a 2-level $(n^{1/2}, n^{1/4})$ partitioning scheme is used and $O(n^{1/2})$ witnesses's exponents are precomputed. Times are given in microseconds.

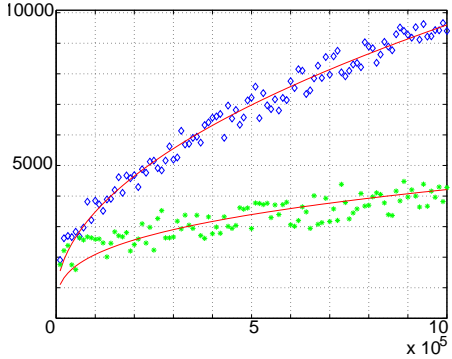
(a) insertion time, in microseconds;

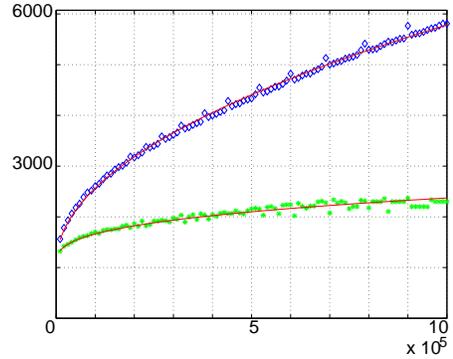
(b) deletion time, in microseconds;

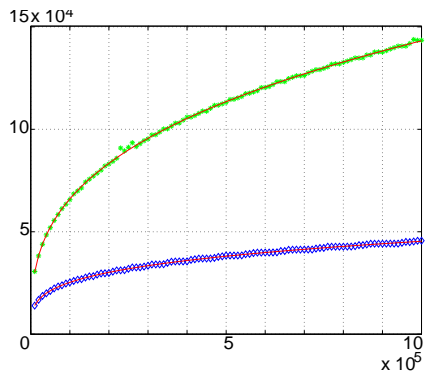
(c) query time, in microseconds.

Figure 5: Insertion and deletion times at the source and query time at the directory for two variants of the hierarchical accumulations approach (Section 5) on dictionaries with up to one million elements. The time for computing the prime representative of an element has been omitted from the insertion time. The stars represent a 2-level ($n^{1/2}$, $n^{1/4}$) partitioning scheme and the diamonds represent a 2-level ($n^{2/3}$, $n^{1/3}$) partitioning scheme.

## 8 Discussion and Conclusion

We have shown how to make the RSA accumulator function the basis for a practical and efficient scheme for authenticated dictionaries, which relies on reasonable cryptographic assumptions similar to those that justify RSA encryption. A distinctive advantage of our approach is that the validation of a query result performed by the user takes constant time and requires computations (a single exponentiation and digital signature verification) simple enough to be performed in devices with very limited computing power, such as a smart card or a mobile phone.

An important aspect of our scheme is that it is dynamic and distributed, thus supporting efficient updates and balancing the work between the source and the directories. A first variation of our scheme achieves a complete tradeoff between the cost of updates at the source and of queries at the directories, with updates taking $O(p + \log(n/p))$ time and queries taking $O(n/p)$ time, for any fixed integer parameter $1 \leq p \leq n$. For example, we can achieve $O(\sqrt{n})$ time for both updates and queries. A second variation of our scheme, suitable for large data sets, achieves $O(n^\epsilon)$-time performance for updates and queries, while keeping $O(1)$ verification time, where $\epsilon > 0$ is any fixed constant.

Our scheme can be easily adapted to contexts, such as certificate revocation queries, where one needs to also validate that an item $e$ is *not* in the set $S$. In this case, we use the standard method of storing in the dictionary not the items themselves, but instead the ranges $r_i = [e_i, e_{i+1}]$ in a sorted list of the elements of $S$ (see, e.g., Kocher [30]). A query for an element $e$ returns a range $r_i = [e_i, e_{i+1}]$ such that $e_i \leq e \leq e_{i+1}$ plus a cryptographic validation of range $r_i$. If $e$ is one of the endpoints of $r_i$, then $e$ in $S$; else ($e_i < e < e_{i+1}$), $e$ is not in $S$. Note that this approach also requires that we have a way of representing some notion of $-\infty$ and $+\infty$. Even so, the overhead adds only a constant factor to all the running times for updates, queries, and validations.

## Acknowledgments

We would like to thank Andrew Schwerin, Giuseppe Ateniese, and Douglas Maughan for several helpful discussions and e-mail exchanges relating to the topics of this paper. The results of this paper were reported in preliminary form in [25].